\documentclass[aps,showpacs,showkeys,amsmath,amssymb,twocolumn,prl,
floatfix,superscriptaddress]{revtex4}

\usepackage{graphicx}
\usepackage{amssymb}
\usepackage{dcolumn}
\usepackage{bm}
\begin{document}
\title{Plasma damping effects on the radiative energy loss of relativistic 
particles}

\author{M.~Bluhm}
\affiliation{SUBATECH, UMR 6457, Universit\'{e} de Nantes, 
Ecole des Mines de Nantes, IN2P3/CNRS. 4 rue Alfred Kastler, 
44307 Nantes cedex 3, France}
\author{P.~B.~Gossiaux}
\affiliation{SUBATECH, UMR 6457, Universit\'{e} de Nantes, 
Ecole des Mines de Nantes, IN2P3/CNRS. 4 rue Alfred Kastler, 
44307 Nantes cedex 3, France}
\author{J.~Aichelin}
\affiliation{SUBATECH, UMR 6457, Universit\'{e} de Nantes, 
Ecole des Mines de Nantes, IN2P3/CNRS. 4 rue Alfred Kastler, 
44307 Nantes cedex 3, France}

\date{\today} 


\begin{abstract}
The energy loss of a relativistic charge undergoing 
multiple scatterings while traversing an infinite, polarizable and 
absorptive plasma is investigated. Polarization and absorption mechanisms 
in the medium are phenomenologically modelled by a complex index of 
refraction. Apart from the known Ter-Mikaelian effect related to 
the dielectric polarization of matter, we find an additional, substantial 
reduction of the energy loss due to the damping of radiation. The observed 
effect is more prominent for larger damping and/or larger energy of the 
charge. A conceivable analog of this phenomenon in QCD could influence 
the study of jet quenching phenomena in ultra-relativistic heavy-ion 
collisions at RHIC and LHC. 
\end{abstract}

\maketitle

The strong suppression of the yields of hadrons with high transverse momentum observed 
in relativistic nuclear collisions~\cite{Adcox02,Adler02} has been interpreted 
as a signature for the formation of a strongly interacting, deconfined and 
dense quark-gluon plasma (QGP)~\cite{Gyulassy05,Shuryak05}, in which energetic 
partons suffer radiative and collisional energy 
loss~\cite{Gyulassy94,Baier97,Baier00,Thoma91n,Braaten91,Thoma91}. Important 
in the context of radiative energy loss of relativistic particles, as realized 
by Landau, Pomeranchuk~\cite{Landau53} and Migdal~\cite{Migdal56} for QED, 
is the possibility of a 
destructive interference between radiation amplitudes when the charged particle 
undergoes multiple scatterings within the formation time of radiation, resulting 
in a suppression of the radiation spectrum compared to the sum of incoherent 
emissions at successive scatterings (LPM effect). This was later on generalized 
to QCD~\cite{Baier97,Baier95}. 

As pointed out by Ter-Mikaelian~\cite{Ter-Mikaelian}, the radiation 
spectrum, and hence the energy loss, is also modified by the dielectric 
polarization of the medium, which gives rise to medium-modifications in the 
dispersion relation of radiated quanta (TM effect). The QCD analog of the TM 
effect was studied in~\cite{Kampfer00,Djordjevic03} by investigating the 
gluon radiation spectrum in the QGP. These approaches made use of dielectric 
functions either including a constant thermal gluon mass~\cite{Kampfer00} or 
being related to the hard thermal loop (HTL) gluon self-energy~\cite{Djordjevic03}. 
In none of these studies, however, the possible additional influence of the 
damping of radiation in an absorptive medium was taken into account. The intent 
of our Letter is to investigate this influence. 

We study in linear response theory the energy loss of an 
energetic point-charge in an absorptive, dielectric medium. Polarization and 
damping of radiation effects are both taken into account by employing a complex 
medium index of refraction. Following the original approach in~\cite{Landau53}, 
the velocity vector $\vec{v}(t)$ of the charge is modelled to change with time due to 
successive scatterings in an infinite medium. Our studies, however, also 
qualitatively apply in the case, in which the size of the medium is large compared 
to the formation length of radiated quanta. A detailed derivation and discussion 
will be reported elsewhere. Here, we want to focus on two essential 
results: (i) Damping of radiation in an absorptive medium can lead to a 
substantial {\it reduction} of the radiative energy loss and (ii) the observed 
effect intensifies with increasing medium damping and/or increasing energy $E$ of 
the charge. Our investigations, being strictly valid for electro-magnetic plasmas, 
represent a classical, abelian approximation for the dynamics of a color charge 
in the QGP. It may, thus, be conceivable that the damping of radiation is also of some impact in 
parton energy loss studies. Throughout this work natural units are used, 
i.~e.~$\hbar=c=1$. 

In line with~\cite{Thoma91}, we determine the energy loss from the negative 
mechanical work $W$ performed on the charge by its electric field. As $W$ accounts 
for the total energy loss of the charge, it incorporates in a finite absorptive medium in 
particular both, the energy radiated out of and the amount of energy dissipated 
inside the medium. Thus, a study of $W$ is particularly suitable for our purposes. 
It is comfortably evaluated in the mixed spatial coordinate and frequency 
representation of its integrand via 
\begin{equation}
\label{equ:Wmech2}
 W = 2 \,Re \left(\int d^3 \vec{r}\,' \int_0^\infty d\omega \, 
 \vec{E}(\vec{r}\,',\omega) \, \vec{j}(\vec{r}\,',\omega)^* \right) ,
\end{equation}
where $\vec{j}(\vec{r}\,',t)=q\vec{v}(t)\delta^{(3)}(\vec{r}\,'-\vec{r}(t))$ is 
the classical current of charge $q$ in space-time coordinates and $\vec{E}$ is 
the total electric field of the charge inside the medium. 

From Maxwell's equations for a linear dispersive medium, one obtains in Fourier-space 
\begin{eqnarray}
\nonumber 
 \vec{E}_{\vec{k}}(\omega) & = & 
 \frac{iq}{(2\pi)^2} 
 \int dt' \frac{e^{i\omega t' -i\vec{k}\vec{r}(t')}}{\omega\epsilon(\omega)} 
 \\ 
\label{equ:DefinEfield}
 & & \hspace{-1.3cm} \times 
 \left\{\frac{\vec{v}_L(t') k^2}{(\omega^2\mu(\omega)\epsilon(\omega)-k^2)}-
 \frac{\vec{v}(t')\omega^2\mu(\omega)\epsilon(\omega)}{(\omega^2\mu(\omega)\epsilon(\omega)-
 k^2)}
 \right\} ,
\end{eqnarray}
where $\vec{v}_L=(\vec{k}\,\vec{v})\vec{k}/k^2$ is the longitudinal component of $\vec{v}$ 
with respect to the wave vector $\vec{k}$, and $\epsilon(\omega)$ and $\mu(\omega)$ denote 
permittivity and permeability of the matter, respectively, which are defined 
to be complex in order to account for the damping of (time-like) excitations in the 
plasma~\cite{Ichimaru}. For positive $\omega$, one finds from~(\ref{equ:Wmech2}) in 
differential form 
\begin{eqnarray}
\nonumber 
 \frac{dW}{d\omega} = 
 Re \Bigg(-\frac{iq^2}{8\pi^4}\int dt \int dt' \int d^3\vec{k} \, 
 \frac{e^{-i\omega(t-t')+i\vec{k}\vec{\Delta r}}}{\omega\epsilon(\omega)} 
 \\
\label{equ:Wmech3}
 \times\left\{\frac{(\vec{k}\,\vec{v}(t))(\vec{k}\,\vec{v}(t'))}{(k^2-
 \omega^2 n^2(\omega))}-
 \frac{\vec{v}(t)\,\vec{v}(t') \omega^2 n^2(\omega)}{(k^2-\omega^2 
 n^2(\omega))}\right\}
  \Bigg) ,
\end{eqnarray}
where $\vec{\Delta r}=\vec{r}(t)-\vec{r}(t')$ and $n^2(\omega)=\mu(\omega)\epsilon(\omega)$ 
is the complex squared index of refraction. 

In the following, we assume for simplicity $\epsilon(\omega)$ and $\mu(\omega)$ to 
depend on $\omega$ only, i.~e.~spatial distortions in the plasma are not considered. 
Then, $dW/d\omega$ in~(\ref{equ:Wmech3}) is sensitive to simple poles in the complex 
momentum-plane and the momentum-integrals in~(\ref{equ:Wmech3}) can easily be 
evaluated analytically by contour integration. The inclusion of an explicit 
$\vec{k}$-dependence in $\epsilon$ and $\mu$ is left for future studies. 

By decomposing the index of refraction into real and imaginary parts, 
$n(\omega)=n_r(\omega)+in_i(\omega)$, and defining 
$\vec{g}=\omega n(\omega) \vec{\Delta r}$ one obtains from~(\ref{equ:Wmech3}) 
\begin{equation}
\label{equ:Wmech5}
 \frac{dW}{d\omega} = Re \left(
 \frac{iq^2}{4\pi^2} \int dt \int dt' \, 
 \frac{\omega^2 n^3(\omega)}{\epsilon(\omega)} 
 e^{-i\omega(t-t')} \mathcal{A}(t,t') 
 \right) 
\end{equation}
with 
\begin{equation}
\label{equ:newmathcalA}
 \mathcal{A}(t,t') = \left[\vec{v}(t)\vec{v}(t')+
 (\nabla_{\vec{g}}\,\vec{v}(t))(\nabla_{\vec{g}}\,\vec{v}(t'))
 \right]\frac{e^{i sgn(n_i)g}}{g} \,.
\end{equation} 
This is the main result of our work. Essential here is the exponential factor 
\begin{equation}
\label{equ:expdampfac}
 e^{i sgn(n_i)g}=e^{i sgn(n_i)\omega n_r\Delta r}e^{-\omega|n_i|\Delta r} , 
\end{equation}
which implies that irrespective of the sign of $n_i(\omega)$, $sgn(n_i)$, the 
mechanical work is exponentially damped for $|n_i|\ne 0$. 
This is a direct consequence of the fact that, depending on $sgn(n_i)$, only one 
of the two simple poles in~(\ref{equ:Wmech3}) contributes to the energy loss. 
Only the phase factor in~(\ref{equ:expdampfac}), associated with $n_r(\omega)$, 
is affected by $sgn(n_i)$. 

Because of the symmetry property of $\mathcal{A}(t,t')$ under variable exchange, 
(\ref{equ:Wmech5}) can be written in a form such that only $t>t'$ has to be 
considered in the time-integration. Omitting as in~\cite{Landau53} those terms 
stemming from the action of $\nabla_{\vec{g}}$ on $1/g$ in~(\ref{equ:newmathcalA}), 
$\mathcal{A}(t,t')$ entering~(\ref{equ:Wmech5}) reduces to 
\begin{equation}
\label{equ:leadingorder}
 \mathcal{A}_{(0)}(t,t')=\frac{1}{g}\left(
 \vec{v}(t)\vec{v}(t')-\frac{(\vec{v}(t)\vec{g})(\vec{v}(t')\vec{g})}{g^2}
 \right) e^{i sgn(n_i)g} \,.
\end{equation} 
For constant $\vec{v}$, $\mathcal{A}_{(0)}(t,t')\equiv 0$ and the corresponding 
energy loss determined via~(\ref{equ:Wmech5}) vanishes. 

Going beyond constant $\vec{v}$, we study, as in~\cite{Landau53}, the 
case, where the velocity $\vec{v}(t')=v'\hat{z}$ is changed due to multiple 
scatterings according to $\vec{v}(t'+\tilde{t})=v'\hat{z}\cos \theta_{\tilde{t}}+
v'\vec{e}_\perp\sin\theta_{\tilde{t}}$ for $\tilde{t}>0$, i.~e.~where $v^2$ 
remains constant. Assuming the relative deflection angle, $\theta_{\bar{t}}$, to 
be small within $\bar{t}=t-t'$, one obtains 
$\vec{v}(t)\vec{v}(t')\simeq v'^2(1-\theta_{\bar{t}}^2/2)$ 
in~(\ref{equ:leadingorder}). Likewise, by omitting terms of order 
$\mathcal{O}(\theta^4)$, one finds 
$\vec{\Delta r}\simeq v'\hat{z}\bar{t}-\frac12 v'\hat{z}\mathcal{I}_2+
v'\vec{e}_\perp\mathcal{I}_1$ with 
$\mathcal{I}_2 = \int_0^{\bar{t}}\theta_\tau^2 d\tau$ 
and $\mathcal{I}_1 = \int_0^{\bar{t}}\theta_\tau d\tau$. Averaging over the 
deflection angles, one finds 
$\langle\Delta r\rangle \simeq v'\bar{t}\sqrt{1-\hat{q}\bar{t}/(3E^2)}$, where the 
parameter $\hat{q} = E^2\langle\theta_{\bar{t}}^2\rangle/(2\bar{t}\,)$ denotes 
one half of the mean accumulated transverse momentum squared of 
the deflected charge per unit time. Approximating $dz=v'dt'$ and 
using~(\ref{equ:leadingorder}), one obtains from~(\ref{equ:Wmech5}) 
\begin{eqnarray}
\nonumber
 \frac{d^2W}{dz d\omega} & \simeq & 
 - Re \Bigg(\frac{2i\alpha}{3\pi} \frac{\hat{q}}{E^2}
 \int_{0}^\infty d\bar{t} \, 
 \frac{\omega n^2}{\epsilon} \cos (\omega \bar{t}\,) 
 \\ \nonumber 
 & & \hspace{4mm}\times\exp\left[i sgn(n_i)\omega n_r \beta \bar{t}
 \left(1-\frac{\hat{q}}{6E^2}\bar{t}\right)\right]
 \\ 
\label{equ:Wmech6}
 & & \hspace{4mm}\times\exp\left[-\omega|n_i|\beta \bar{t}
 \left(1-\frac{\hat{q}}{6E^2}\bar{t}\right)\right] 
 \Bigg) 
\end{eqnarray} 
with $\beta = v'$ and coupling $\alpha=q^2/(4\pi)$. 

For $\hat{q}=0$, i.~e.~when the charge suffers no deflections, 
$d^2W/(dz d\omega)$ vanishes. Furthermore, in the vacuum limit, i.~e.~when setting 
$\epsilon(\omega)=\mu(\omega)=1$,~(\ref{equ:Wmech6}) becomes the negative of the 
radiation intensity determined in~\cite{Landau53}. Thus, we interpret the negative of 
expression~(\ref{equ:Wmech6}) as radiative energy loss spectrum per unit 
length. We restrict ourselves to the case $sgn(n_i)=sgn(n_r)$ in order to account 
for an actual loss of energy. Moreover, we do not distinct in the following between 
longitudinal and transverse excitations by setting $\mu(\omega)=1$, which implies 
$\epsilon_L=\epsilon_T=\epsilon$ for isotropic and homogeneous media~\cite{Ichimaru}. 
A differentiation between longitudinal and transverse excitations is left for future 
investigations. 

As the approach assumes small deflection angles for arbitrary values of $\bar{t}$, 
it is necessary to impose as physical constraint a natural upper boundary in the 
$\bar{t}$-integral. We restrict~(\ref{equ:Wmech6}) to $\bar{t}\leq 2E^2/\hat{q}$, 
i.~e.~exactly where $\langle\Delta r\rangle/v'$ reaches its maximum. Increasing 
the cut-off value reasonably does not influence our numerical results presented 
below. 

We use for the complex squared index of refraction $n^2(\omega)\equiv\epsilon(\omega)$ 
the following formal ansatz 
\begin{equation}
\label{equ:refrindex}
 n^2(\omega)= 1-\frac{m^2}{\omega^2}+2 i \frac{\Gamma}{\omega} \,. 
\end{equation}
This structure is based on the assumption that radiated quanta formed inside a 
medium follow medium-modified dispersion relations of plasma modes, 
cf.~\cite{Kampfer00,Djordjevic03}, acquiring a finite in-medium mass $m$ and being 
damped in the absorptive medium with a rate related to 
$\Gamma$~\cite{Peshier0405,Pisarski}. Expression~(\ref{equ:refrindex}) is connected 
with a Lorentz-type spectral function for intermediate hard quanta~\cite{Peshier0405}, 
where $m$ and $\Gamma$ are in general free parameters, which may depend on plasma 
temperature, coupling and/or frequency. The corresponding dispersion relation of 
plasma modes follows from $Re(k^2-\omega^2\epsilon(\omega))=0$. This implies (i) the 
absence of radiation for $\omega<m$ (for $\omega\to m^+$, the phase velocity 
squared of the plasma modes diverges and becomes negative for $\omega<m$), 
(ii) time-like plasma modes and 
(iii) $Im(\epsilon)=2\Gamma/\omega\ne 0$ for $\Gamma\ne 0$ with support 
in the $\omega$-region, where the plasma modes exist. Our ansatz for the dielectric 
function, therefore, differs from an $\epsilon(\omega)$ deduced from a leading-order 
HTL self-energy expression, cf.~\cite{Thoma91,Djordjevic03}. In this case, 
plasma modes are also time-like~\cite{leBellac,Weldon82,BlaizotIancu}, while the 
support of $Im(\epsilon)$ is restricted to the space-like region. Only at 
next-to-leading-order, the damping of plasma modes emerges in HTL-based 
approaches. 

\begin{figure}[t]
\centering
\vspace{3mm}
\includegraphics[scale=0.28]{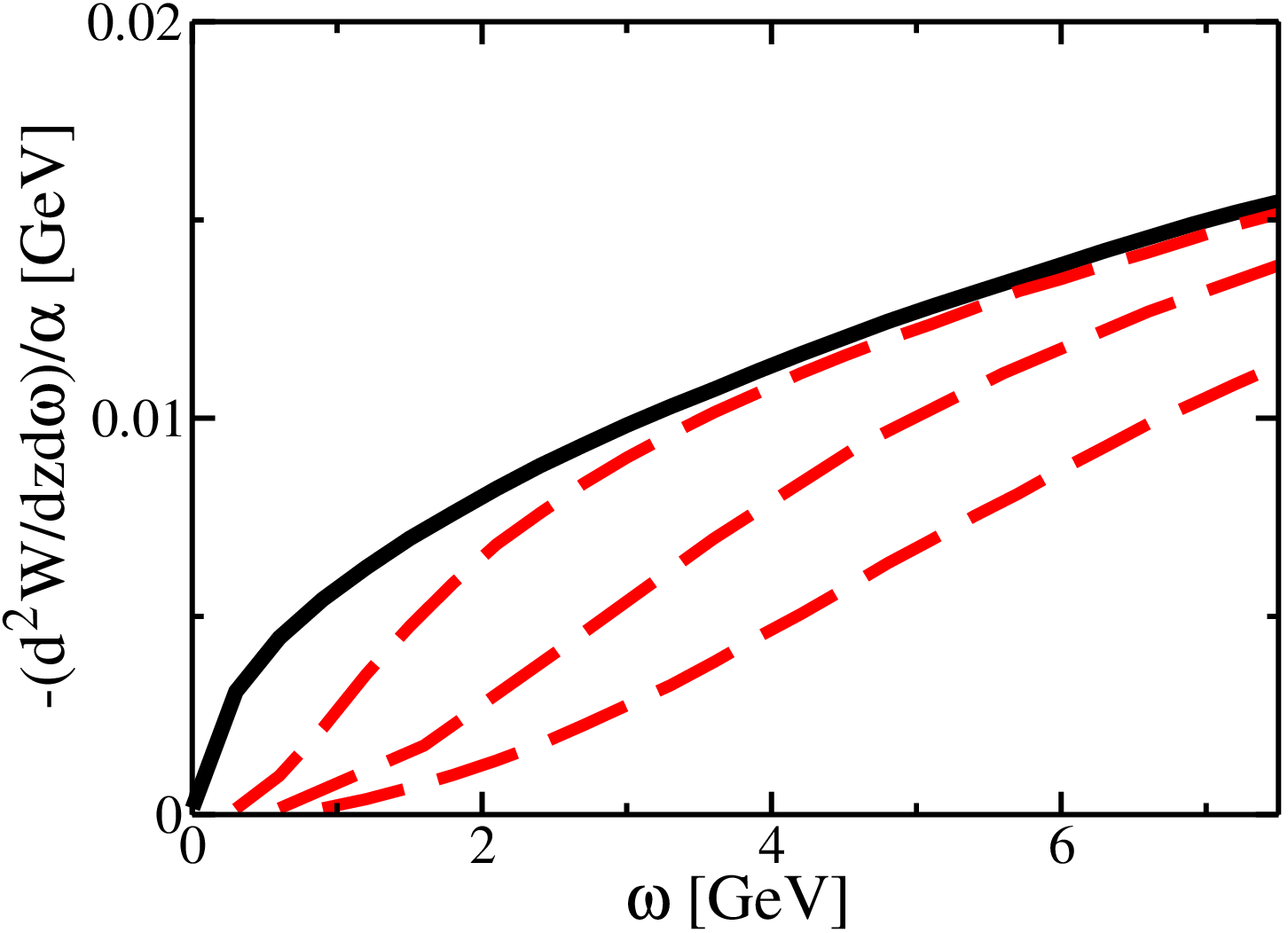} 
\caption[]{\label{Fig:1} (Color online) 
Radiative energy loss spectrum as a function of $\omega$ 
for $\Gamma=0$. The kinematic parameters are chosen as $E=20$~GeV, 
charge mass $M=1$~GeV and $\hat{q}=2.5$~GeV$^2/$fm. Black solid curve 
shows the vacuum limit (setting $m=0$), while the red long-dashed 
curves depict the corresponding results for a medium with real 
$n^2(\omega)$ ($m=0.3,\,0.6,\,0.9$~GeV from top to bottom, 
respectively).}
\end{figure}

We now want to quantify~(\ref{equ:Wmech6}), simplifying however the considerations 
by restricting ourselves to constant $\Gamma$ and $m$ values inspired by~\cite{Cassing07}. 
We discuss first the case of a polarizable, non-damping medium employing 
$n^2(\omega)$ from~(\ref{equ:refrindex}) with $\Gamma=0$. Figure~\ref{Fig:1} exhibits 
the radiative energy loss spectrum for $\Gamma=0$ as a function of 
$\omega$ for $m=0.3,\,0.6,\,0.9$~GeV (red long-dashed curves from top to bottom, 
respectively). For comparison also the vacuum limit ($m=0$) is 
shown (black solid curve). With increasing in-medium mass an 
increasing reduction of the spectrum is observed (TM effect, 
cf.~also~\cite{Kampfer00,Djordjevic03}). This reduction is more pronounced at 
small $\omega$, while the generic approach of the vacuum result with increasing 
$\omega$ is hampered for increasing $m$. 

\begin{figure}[t]
\centering
\vspace{3mm}
\includegraphics[scale=0.28]{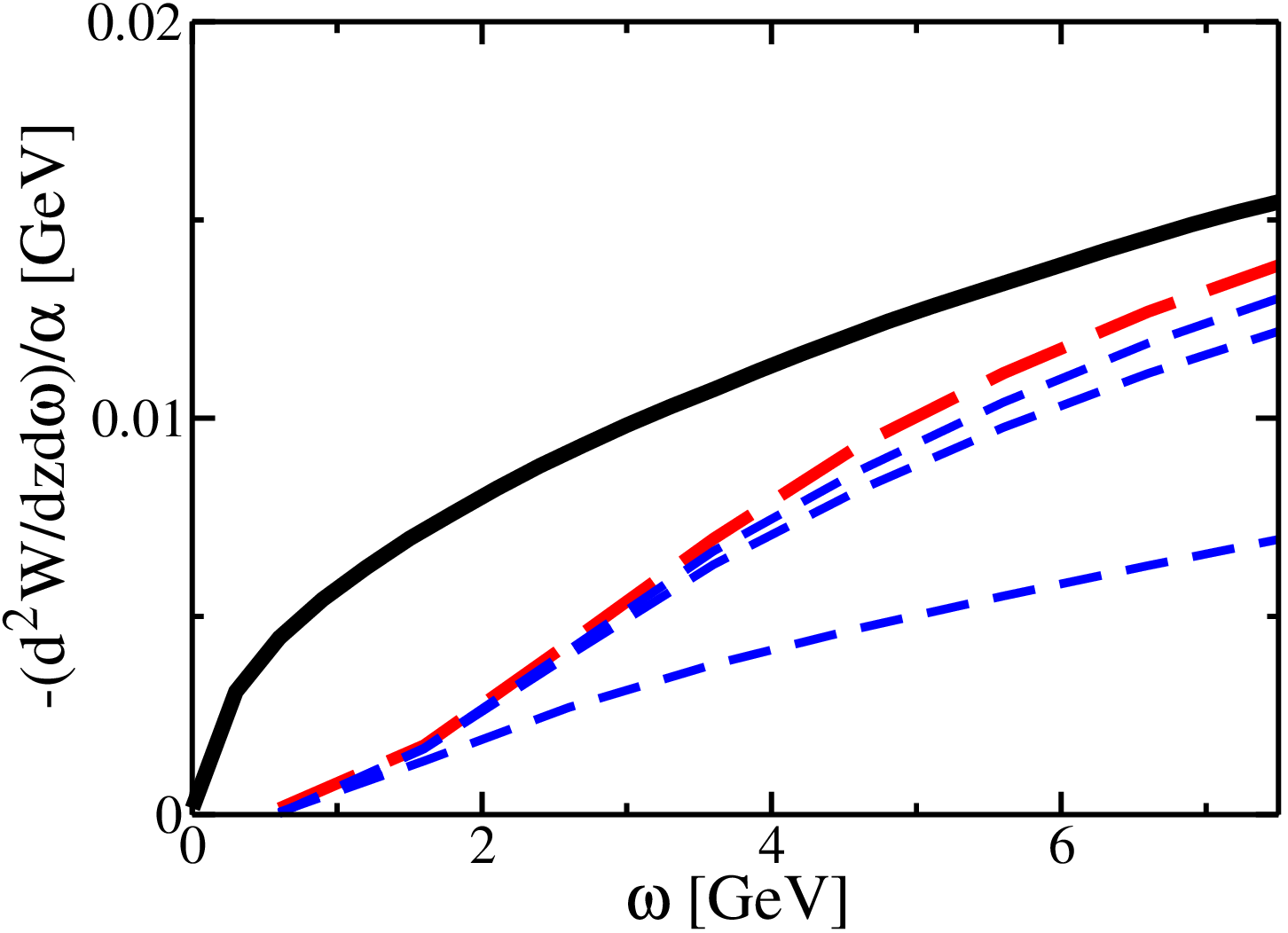} 
\caption[]{\label{Fig:2} (Color online) 
Radiative energy loss spectrum as a function of $\omega$ 
with the same kinematic parameters as in Fig.~\ref{Fig:1}. 
The black solid curve shows the vacuum limit and the red 
long-dashed curve depicts the result for a medium with real 
$n^2(\omega)$ using $m=0.6$~GeV. The blue short-dashed curves 
exhibit the additional influence of the damping of radiation, 
employing $n^2(\omega)$ from~(\ref{equ:refrindex}) 
with $\Gamma=5,\,10,\,50$~MeV (from top to bottom, 
respectively).}
\end{figure} 
The additional effect of the damping of radiation in an absorptive medium 
is shown in Fig.~\ref{Fig:2} for different values of $\Gamma=5,\,10,\,50$~MeV (blue 
short-dashed curves from top to bottom, respectively) and fixed $m=0.6$~GeV. 
Even for small values of $\Gamma$ (the maximal ratio considered in Fig.~\ref{Fig:2} 
is $\Gamma/m=0.083$), damping of radiation leads to a significant reduction of the 
radiative energy loss spectrum. This is a natural consequence of the sensitivity 
of~(\ref{equ:Wmech6}) on the poles in the complex $k$-plane determined from 
$\omega^2-k^2-m^2+2i\Gamma\omega=0$, which implies a non-negligible effect of 
$\Gamma$ even for small $\Gamma/m$. In the limit $\omega\rightarrow m^+$, the 
behavior for a polarizable, non-damping medium is recovered. 

\begin{figure}[t]
\centering
\vspace{3mm}
\includegraphics[scale=0.27]{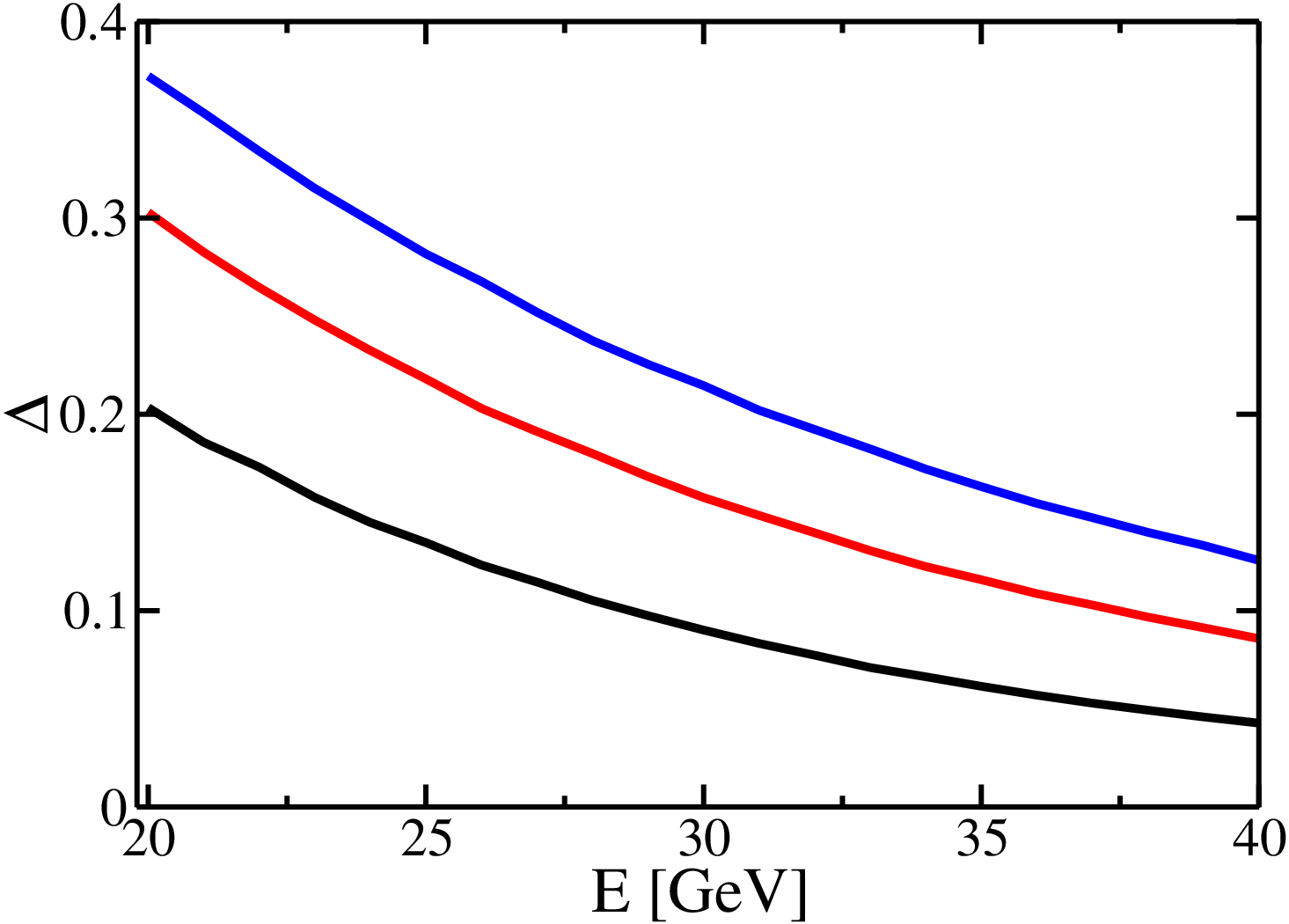} 
\caption[]{\label{Fig:3} (Color online) 
Ratio $\Delta$ as a function of $E$ for constant $\omega=3,\,5,\,7$~GeV 
(solid curves from bottom to top, respectively), and fixed $\Gamma=50$~MeV 
and $m=0$. Charge mass $M$ and parameter $\hat{q}$ as in 
Figs.~\ref{Fig:1} and~\ref{Fig:2}. The ratio $\Delta\to 1$ for negligible 
damping effects.}
\end{figure} 
It is interesting to investigate the dependence 
of~(\ref{equ:Wmech6}) on the energy of the charge for a given $\Gamma$. 
In Fig.~\ref{Fig:3}, the ratio of the radiative energy loss spectrum between 
an absorptive ($\Gamma\ne 0$) and a non-absorptive ($\Gamma=0$) medium, 
$\Delta\equiv d^2W_{\Gamma\ne 0}/d^2W_{\Gamma=0}$, is exhibited 
as a function of $E$ for constant $\omega$, $\Gamma=50$~MeV and $m=0$. 
This allows for studying the relative importance of the damping of radiation 
in an absorptive medium compared to the vacuum. We observe a sensitive energy 
dependence with increasing importance of damping effects for increasing $E$. 

The observations made in Figs.~\ref{Fig:1}-\ref{Fig:3} can be understood by 
qualitatively analyzing~(\ref{equ:Wmech6}). Here, we focus on the 
discussion of observations made in the regime $\omega\gg m\gg\Gamma$. In the case 
of a non-damping but polarizable medium, the vacuum result will be approached 
with increasing $\omega$ as the exponential damping factor in~(\ref{equ:Wmech6}) 
is $1$ and $n_r\to 1$ with increasing $\omega$. 

Analyzing~(\ref{equ:Wmech6}) for $\Gamma\ne 0$, one finds 
from~(\ref{equ:refrindex}) that $\omega|n_i|\to\Gamma$ and 
$sgn(n_i)\omega n_r\to\omega$ in the considered $\omega$-regime. 
In this limit,~(\ref{equ:Wmech6}) gives 
\begin{eqnarray}
\nonumber
 \frac{d^2W}{dz d\omega} & \simeq & 
 \frac{\alpha\omega\hat{q}}{3\pi E^2} 
 \int_{0}^\infty d\bar{t} \exp\left[-\Gamma\beta \bar{t}
 \left(1-\frac{\hat{q}}{6E^2}\bar{t}\right)\right] \\ 
\nonumber 
 & & \hspace{3mm}\times \left\{ 
 \sin\left[\omega \bar{t} (\beta-1) 
 -\omega \beta \frac{\hat{q}}{6E^2}\bar{t}^{\,2}\right] \right. \\
\label{equ:WmechDM2}
 & & \hspace{8mm}\left.
 + \sin\left[\omega \bar{t} (\beta+1) 
 - \omega \beta \frac{\hat{q}}{6E^2}\bar{t}^{\,2}\right]
 \right\} .
\end{eqnarray}
The exponential damping factor in~(\ref{equ:WmechDM2}) is responsible for the 
suppression of the spectrum compared to the $\Gamma=0$ case. It is, unless $\Gamma$ 
depends on $\omega$, formally frequency independent. 

In order to elucidate the impact of the exponential damping factor, it is necessary 
to study the formation time $t_f$ of radiation from a relativistic charge, which 
in general depends on $\omega$. A detailed discussion of $t_f$ in a polarizable and 
absorptive medium is presented in~\cite{Bluhm11}. We determine $t_f$ from the 
phase factor $\Phi(t)$ of the dominant sine-function in~(\ref{equ:WmechDM2}). 
As is well known, radiation can be considered as decoupled from its emitter, once a 
phase $\Phi(t_f)\sim 1$ has been accumulated. This leads to the condition 
$1\sim \omega t_f(1-\beta)+\omega\beta\,\hat{q}\,t_f^2/(6E^2)$. 
Accordingly, $t_f$ is roughly given by the minimum of the two limiting solutions, 
$2E^2/(\omega M^2)$ and $E\sqrt{6/(\omega\hat{q})}$. 
The increase of $t_f$ with $E$ explains why a rather small $\Gamma$ can lead to 
a large suppression of radiation: 
The exponential damping factor in~(\ref{equ:WmechDM2}) reduces from $1$ at $\bar{t}=0$ to 
approximately $\exp[-\Gamma t_f]$ within the formation time 
interval. This gives rise to the observed sensitive $\Gamma$- and 
$E$-dependence of the spectrum, unless $t_f$ is comparable to or even larger than another 
competing time scale $t_d\sim 1/\Gamma$, cf.~\cite{Bluhm11}. Then, the 
exponential damping factor becomes of order $\mathcal{O}(1/e)$, such that the 
amplitude in~(\ref{equ:WmechDM2}) is damped away before radiation could be 
formed. 

In summary, we have shown that the radiative energy loss of an energetic 
charge can be substantially reduced in an absorptive medium (modelled by 
$n_i\ne 0$). This occurs in addition to the known TM effect, which already 
leads to a reduction of the radiative energy loss in a polarizable but 
non-absorptive medium (described by $n_r\ne 0$ and $n_i=0$). 
The observed effect increases with increasing medium damping 
and/or increasing energy of the charge. Our investigations, being restricted 
here to the regime $\omega\ll E$, represent a classical, non-abelian 
approximation for the dynamics of a color charge in the QGP. 
Important effects specific to QCD, such as gluon rescatterings, 
are, however, still missing. This will be explored in future work. 

We gratefully acknowledge valuable and insightful discussions with 
B.~K\"ampfer, T.~Gousset and S.~Peign\'{e}. The work is supported by 
the European Network I3-HP2 Toric and the ANR research program 
``Hadrons@LHC'' (grant ANR-08-BLAN-0093-02). 

\end{document}